\documentclass[11pt]{article}
\usepackage{amssymb,amsmath,amsfonts}
\usepackage{graphicx}
\usepackage{graphics}
\usepackage{eepic,epsfig}

 

\begin{document}

\title{Vacuum densities for a brane intersecting the AdS boundary}
\author{E. R. Bezerra de Mello$^{1}$\thanks{
E-mail: emello@fisica.ufpb.br}, A. A. Saharian$^{2,1}$\thanks{%
E-mail: saharian@ysu.am}, M.R. Setare$^{3}$\thanks{%
E-mail: rezakord@ipm.ir} \vspace{0.3cm} \\
%EndAName
\textit{$^{1}$Departamento de F\'{\i}sica, Universidade Federal da Para\'{\i}%
ba}\\
\textit{58.059-970, Caixa Postal 5.008, Jo\~{a}o Pessoa, PB, Brazil}\vspace{%
0.3cm}\\
\textit{$^2$Department of Physics, Yerevan State University,}\\
\textit{1 Alex Manoogian Street, 0025 Yerevan, Armenia} \vspace{0.3cm}\\
\textit{$^{3}$Department of Science, Campus of Bijar,}\\
\textit{University of Kurdistan, Bijar, Iran}}
\maketitle

\begin{abstract}
We investigate the Wightman function, the bulk-to-boundary propagator, the
mean field squared and the vacuum expectation values of energy-momentum
tensor for a scalar field in AdS spacetime, in the presence of a brane
perpendicular to the AdS boundary. On the brane the field operator obeys
Robin boundary condition. The vacuum expectation values are decomposed into
the boundary-free AdS and brane-induced contributions. In this way, for
points away from the brane, the renormalization is reduced to the one in
pure AdS spacetime. It is shown that at proper distances from the brane
larger than the AdS curvature radius the brane-induced expectation values
decay as power-law for both massless and massive scalars. This behavior is
in contrast to that for a plane boundary in Minkowski spacetime, with an
exponential decay for massive fields. For Robin boundary conditions
different from Dirichlet and Neumann ones, the brane-induced part in the
energy density is positive near the brane and negative at large distances.
For Dirichlet/Neumann boundary condition the corresponding energy density is
negative/positive everywhere. We show that, for a fixed value of the proper
distance from the brane, near the AdS boundary, the Neumann boundary
condition is an "attractor" in the general class of Robin boundary
conditions, whereas Dirichlet boundary condition is an "attractor" near the
horizon.
\end{abstract}

\bigskip


\bigskip

\section{Introduction}

\label{sec:introd}

Anti-de Sitter (AdS) spacetime is among the most popular geometries in
quantum field theory on curved backgrounds. This interest is motivated by
several reasons. First of all, because of its high symmetry, many problems
are exactly solvable on AdS bulk and this may shed light on the influence of
a classical gravitational field on the quantum matter in more general
geometries. The importance of AdS spacetime as a gravitational background
increased by its natural appearance as a stable ground state solution in
extended supergravity and in string theories. The AdS geometry plays a
crucial role in two exciting developments in theoretical physics of the last
20 years such as the AdS/CFT correspondence and the braneworld scenario. The
first one, the AdS/CFT correspondence \cite{Mald98} (see \cite{Ahar00} for a
review), represents a realization of the holographic principle and relates
string theories or supergravity in the AdS bulk with a conformal field
theory living on its boundary. It enables to study conformal field theory
and non-perturbative quantum gravity at the same time. The braneworld
scenario (for reviews on braneworld gravity see \cite{Ruba01}) offers a new
perspective on the hierarchy problem between the gravitational and
electroweak mass scales. In the corresponding models, our world is
represented by a sub-manifold, a three-brane, embedded in a higher
dimensional spacetime and the small coupling of four-dimensional gravity is
generated by the large physical volume of extra dimensions.

The investigations of quantum effects both in AdS/CFT and braneworld setups
are of considerable interest in particle physics and in cosmology. An
inherent feature in these setups is the presence of boundaries and the
fields which propagate in the bulk will give Casimir-type contributions to
the vacuum expectation values of physical observables (for reviews of the
Casimir effect see \cite{Eliz94}). In particular, in braneworld scenario,
vacuum forces arise acting on the branes which, depending on the type of a
field and boundary conditions imposed, can either stabilize or destabilize
the braneworld. The Casimir energy gives a contribution to both the brane
and bulk cosmological constants and, hence, has to be taken into account in
the self-consistent formulation of the corresponding models. Motivated by
these issues, the quantum vacuum effects induced by branes in AdS bulk have
received a great deal of attention. The Casimir energy and the forces for
parallel branes are investigated both for scalar and fermionic fields \cite%
{Fabi00}. Local Casimir densities are discussed in \cite{Knap04}. Quantum
vacuum effects in higher-dimensional generalizations of the AdS spacetime
with compact internal spaces have been studied in \cite{Flac03}. The vacuum
polarization induced by a cosmic string in AdS spacetime is investigated in
\cite{Beze12} for both scalar and fermionic fields.

In the most of the papers cited above the branes are considered to be
parallel to the AdS boundary. Recently, there have been some attempts to
extend AdS/CFT correspondence to the case with boundaries in CFT side \cite%
{Taka11}. In an effective description of the corresponding holographic dual
(AdS/BCFT correspondence) a boundary is introduced in AdS bulk which crosses
the AdS boundary and is anchored at the boundary of CFT. In the construction
of \cite{Taka11}, on the boundary in AdS bulk, the Neumann boundary
condition is imposed in the gravity sector. Another class of problems with
boundaries in the bulk crossing the AdS boundary, recently appeared related
to a geometric procedure for the evaluation of the entanglement entropy in
the context of the AdS/CFT correspondence suggested in \cite{Ryu06} (for an
overview see \cite{Nish09}). In accordance with this procedure, the
entanglement entropy for a bounded region in CFT with respect to its spatial
complement is expressed in terms of the area of the minimal surface in the
bulk, anchored at the boundary of that region. In quantum field theory, the
boundaries in both AdS and CFT will lead to the shifts in the expectation
values of physical quantities describing the properties of the vacuum. These
effects should be taken into account in discussions of the stability of the
corresponding models.

In the present paper, for a scalar quantum field with general curvature
coupling parameter, we consider an exactly solvable problem with a flat
brane in AdS spacetime perpendicular to its boundary. This model is a
holographic dual of BCFT defined on a half-space. In order to clarify the
role of the boundary condition, we impose on the field operator a general
Robin condition. Our main interest will be the changes in the properties of
the quantum vacuum induced by the presence of the brane. The important
quantities that characterize the local properties of the vacuum are the
expectation values of the field squared and energy-momentum tensor. The
latter serves as a source in the right-hand side of semiclassical Einstein
equations and plays an important role in considerations of the back-reaction
from quantum effects.

The organization of the paper is as follows. In the next section we evaluate
the positive-frequency Wightman function and the bulk-to-boundary propagator. 
The corresponding expressions are
explicitly decomposed into the boundary-free and brane-induced
contributions. On the base of this, in sections \ref{sec:phi2} and \ref%
{sec:EMT} we investigate the mean field squared and the vacuum expectation
value of the energy-momentum tensor. Various asymptotics for the
brane-induced contributions are discussed and the corresponding results are
compared with those for a Robin plate in Minkowski spacetime. Section \ref%
{sec:Conc} summarizes the main results of the paper.

\section{Two-point functions}

\label{sec:WF}

Let us consider a scalar field $\varphi (x)$ on background of a $(D+1)$%
-dimensional AdS spacetime with the curvature radius $\alpha $. The
corresponding line element will be taken in the form
\begin{equation}
ds^{2}=g_{ik}dx^{i}dx^{k}=e^{-2y/\alpha }\eta _{\mu \nu }dx^{\mu }dx^{\nu
}-dy^{2},  \label{metric}
\end{equation}%
where $\eta _{\mu \nu }=\mathrm{diag}(1,-1,\ldots ,-1)$ is the metric tensor
for the $D$-dimensional Minkowski spacetime, $i,k=0,1,\ldots ,D$, and $\mu
,\nu =0,1,\ldots ,D-1$. For a field with the curvature coupling parameter $%
\xi $ the field equation has the form
\begin{equation}
(g^{ik}\nabla _{i}\nabla _{k}+m^{2}+\xi R)\varphi (x)=0,  \label{fieldeq}
\end{equation}%
where $\nabla _{i}$ is the covariant derivative operator and $R$ is the
Ricci scalar. The latter is related to the curvature radius as $%
R=-D(D+1)/\alpha ^{2}$. For special cases of minimally and conformally
coupled scalars one has $\xi =0$ and $\xi =\xi _{D}=(D-1)/(4D)$,
respectively. By a coordinate transformation $z=\alpha e^{y/\alpha }$ the
line element (\ref{metric}) is written in a conformally-flat form $%
ds^{2}=(\alpha /z)^{2}\eta _{ik}dx^{i}dx^{k}$ with $x^{D}=z$ and with the
conformal factor $(\alpha /z)^{2}$. In terms of the coordinate $z$, the AdS
boundary and the horizon are presented by the hypersurfaces $z=0$ and $%
z=\infty $, respectively.

Our main interest in this paper are the vacuum expectation values (VEVs) of
the field squared and of the energy-momentum tensor in the presence of a
flat brane at $x^{1}=0$. In what follows, for definiteness, we shall
consider the region $x^{1}\geqslant 0$. The boundary-induced contributions
in the VEVs of the field squared and of the diagonal components of the
energy-momentum tensor are symmetric under the reflection $x^{1}\rightarrow
-x^{1}$, whereas the off-diagonal component $\langle T_{D}^{1}\rangle _{b}$
(see below) changes the sign. On the brane we impose the Robin boundary
condition
\begin{equation}
\left( 1+\beta \partial _{1}\right) \varphi (x)=0,\quad x^{1}=0,
\label{boundcond}
\end{equation}%
with a constant coefficient $\beta $. The corresponding results for
Dirichlet and Neumann boundary conditions are obtained as special cases
corresponding to $\beta =0$ and $\beta =\infty $. The geometry under
consideration presents a holographic description of a BCFT living on the
hypersurface $z=0$, $x^{1}\geqslant 0$. The Robin boundary condition
naturally arises for scalar bulk fields in braneworld models. The parameter $%
\beta $ encodes the properties of the brane. For example, in the
Randall-Sundrum braneworld models with the branes parallel to the AdS
boundary, this coefficient is expressed in terms of the curvature coupling
parameter and brane mass terms for a scalar field \cite{Gher00}.

\subsection{Wightman function}

The imposition of the boundary condition modifies the spectrum for the
vacuum fluctuations of quantum fields. As a consequence, the VEVs of
physical quantities are shifted with respect to the VEVs in the
boundary-free geometry. The renormalized VEVs of bilinear combinations of
the field operator are obtained from the two-point functions after an
appropriate renormalization procedure. In this section, we shall evaluate
the positive-frequency Wightman function,
\begin{equation}
W(x,x^{\prime })=\langle 0|\varphi (x)\varphi (x^{\prime })|0\rangle ,
\label{Wightfunc}
\end{equation}%
where $|0\rangle $ stands for the vacuum state. This function also
determines the response of Unruh-DeWitt-type particle detectors interacting
locally with a quantum field under consideration (see, for instance, \cite%
{Birr82}). For the evaluation of the Wightman function we shall employ the
direct summation approach over a complete orthonormal set of mode functions $%
\left\{ \varphi _{\sigma }(x),\varphi _{\sigma }^{\ast }(x)\right\} $,
specified by a set of quantum numbers $\sigma $ and obeying the boundary
condition (\ref{boundcond}). The corresponding mode-sum formula reads
\begin{equation}
W(x,x^{\prime })=\sum_{\sigma }\varphi _{\sigma }(x)\varphi _{\sigma }^{\ast
}(x^{\prime }),  \label{Wightvev}
\end{equation}%
where $\sum_{\sigma }$ includes summation over discrete quantum numbers and
the integration over continuous ones.

In the problem under consideration, the mode functions can be presented in
the factorized form
\begin{equation}
\varphi _{\sigma }(x)=Cz^{D/2}J_{\nu }(\gamma z)\cos [\lambda x^{1}+\alpha
_{0}(\lambda )]e^{i\mathbf{kx}-i\omega t},  \label{eigfunc1}
\end{equation}%
where $\mathbf{x}=(x^{2},\ldots ,x^{D-1})$, $\mathbf{k}=(k_{2},\ldots
,k_{D-1})$, $k=|\mathbf{k}|$, and%
\begin{equation}
\omega =\sqrt{k^{2}+\lambda ^{2}+\gamma ^{2}}.  \label{om}
\end{equation}%
In (\ref{eigfunc1}), $J_{\nu }(x)$ is the Bessel function with the order
\begin{equation}
\nu =\sqrt{(D/2)^{2}-D(D+1)\xi +m^{2}\alpha ^{2}}.  \label{nu}
\end{equation}%
For a conformally coupled massless scalar one has $\nu =1/2$ and $J_{\nu
}(x)=\sqrt{\pi /2x}\sin x$. For imaginary $\nu $ the ground state becomes
unstable \cite{Brei82} and, in what follows, we shall assume that this
parameter is real. Note that in defining the modes (\ref{eigfunc1}) we have
imposed Dirichlet boundary condition on the AdS boundary.

From the boundary condition (\ref{boundcond}) for the function $\alpha
_{0}(\lambda )$ one finds%
\begin{equation}
e^{2i\alpha _{0}(\lambda )}=\frac{i\beta \lambda -1}{i\beta \lambda +1}.
\label{alf0}
\end{equation}%
For $\beta >0$, in addition to the modes (\ref{eigfunc1}), there is a mode
with the dependence on the coordinate $x^{1}$ in the form $e^{-x^{1}/\beta }$
which describes a bound state. For this mode one has $\omega =\sqrt{%
k^{2}+\gamma ^{2}-1/\beta ^{2}}$ and there is a region in the space $(\gamma
,k)$ where the energy of the mode becomes imaginary. This signals about the
instability of the vacuum. Here the situation is essentially different from
that for a plate in Minkowski spacetime. In the latter geometry, for a
massive scalar field, under the condition $\beta >1/m$, the bound state has
a positive energy and the vacuum is stable. In the discussion below we shall
assume that $\beta <0$.

The coefficient $C$ in Eq. (\ref{eigfunc1}) is determined from the
orthonormality condition%
\begin{equation}
\int d^{D}x\,\sqrt{|g|}g^{00}\varphi _{\sigma }(x)\varphi _{\sigma ^{\prime
}}^{\ast }(x)=\frac{\delta _{\sigma \sigma ^{\prime }}}{2\omega },
\label{NormCond}
\end{equation}%
and is given by the expression%
\begin{equation}
|C|^{2}=\frac{2\gamma }{\left( 2\pi \alpha \right) ^{D-1}\omega }.
\label{Cnorm}
\end{equation}%
In (\ref{NormCond}), the integration with respect to $x^{1}$ goes over $%
0\leqslant x^{1}<\infty $.

Substituting the eigenfunctions (\ref{eigfunc1}) into the mode sum (\ref%
{Wightvev}), for the Wightman function one finds%
\begin{eqnarray}
W(x,x^{\prime }) &=&W_{0}(x,x^{\prime })+\frac{(zz^{\prime })^{D/2}}{\left(
2\pi \alpha \right) ^{D-1}}\int d\mathbf{k}\int_{0}^{\infty }d\gamma
\int_{0}^{\infty }d\lambda \,\frac{\gamma }{\omega }  \notag \\
&&\times J_{\nu }(\gamma z)J_{\nu }(\gamma z^{\prime })\cos [\lambda
(x^{1}+x^{1\prime })+2\alpha _{0}(\lambda )]e^{i\mathbf{k}\Delta \mathbf{x}%
-i\omega \Delta t}.  \label{W}
\end{eqnarray}%
where $\Delta \mathbf{x=x}-\mathbf{x}^{\prime }$, $\Delta t=t-t^{\prime }$.
Here,%
\begin{equation}
W_{0}(x,x^{\prime })=\frac{(zz^{\prime })^{D/2}}{2(2\pi \alpha )^{D-1}}\int d%
\mathbf{k}_{D-1}\int_{0}^{\infty }d\gamma \,\frac{\gamma }{\omega _{0}}%
J_{\nu }(\gamma z)J_{\nu }(\gamma z^{\prime })e^{i\mathbf{k}_{D-1}\Delta
\mathbf{x}_{D-1}-i\omega _{0}\Delta t},  \label{W0}
\end{equation}%
is the Wightman function in AdS spacetime in the absence of the boundary at $%
x^{1}=0$ (for two-point functions in AdS spacetime see \cite{Burg85,Camp92}%
). In (\ref{W0}), $\mathbf{x}_{D-1}=(x^{1},x^{2},\ldots ,x^{D-1})$, $\mathbf{%
k}_{D-1}=(k_{1},k_{2},\ldots ,k_{D-1})$ and $\omega _{0}=\sqrt{|\mathbf{k}%
_{D-1}|^{2}+\gamma ^{2}}$. The boundary-free Wightman function is expressed
in terms of the hypergeometric function as
\begin{equation}
W_{0}(x,x^{\prime })=\frac{\alpha ^{1-D}f_{\nu }(u_{-})}{2^{D/2+\nu +1}\pi
^{D/2}}\,,  \label{W01}
\end{equation}%
where, for the further convenience, we have introduced the notation
\begin{equation}
f_{\nu }(u)=\frac{\Gamma (\nu +D/2)}{\Gamma (\nu +1)u^{\nu +D/2}}%
\,{}_{2}F_{1}\left( \frac{D+2\nu +2}{4},\frac{D+2\nu }{4};\nu +1;\frac{1}{%
u^{2}}\right) ,  \label{fnu}
\end{equation}%
and%
\begin{equation}
u_{-}=1+[(\Delta z)^{2}+(\Delta \mathbf{x}_{D-1})^{2}-(\Delta
t)^{2}]/(2zz^{\prime }).  \label{u}
\end{equation}%
Note that the quantitiy $u_{-}$ is expressed in terms of the geodesic
distance $\sigma (x,x^{\prime })$ between the points $x$ and $x^{\prime }$
by the relation $u_{-}=\cosh (\sigma (x,x^{\prime })/a)$ for $(\Delta
z)^{2}+(\Delta \mathbf{x}_{D-1})^{2}>(\Delta t)^{2}$ and by $u_{-}=\cos
(\sigma (x,x^{\prime })/a)$ for $(\Delta z)^{2}+(\Delta \mathbf{x}%
_{D-1})^{2}<(\Delta t)^{2}$.

The second term on the right-hand side of Eq. (\ref{W}) is induced by the
brane at $x^{1}=0$. For the further transformation of this part we write%
\begin{equation*}
\cos [\lambda (x^{1}+x^{1\prime })+2\alpha _{0}(\lambda )]=\frac{1}{2}%
\sum_{j=\pm 1}e^{ji\lambda (x^{1}+x^{1\prime })}\frac{i\beta \lambda -j}{%
i\beta \lambda +j},
\end{equation*}%
and rotate the integration contour over $\lambda $ by angle $j\pi /2$ for
the term with $e^{ji\lambda (x^{1}+x^{1\prime })}$. After integration over
the angular part of $\mathbf{k}$, the Wightman function is presented in the
form%
\begin{eqnarray}
W(x,x^{\prime }) &=&W_{0}(u)+\frac{\alpha ^{1-D}(zz^{\prime })^{D/2}}{\left(
2\pi \right) ^{D/2}|\Delta \mathbf{x}|^{D/2-2}}\int_{0}^{\infty
}dk\,k^{D/2-1}J_{D/2-2}(k|\Delta \mathbf{x}|)\,  \notag \\
&&\times \int_{0}^{\infty }d\gamma \,\gamma J_{\nu }(\gamma z)J_{\nu
}(\gamma z^{\prime })\int_{0}^{\infty }dx\,\cosh (\Delta tx)\frac{%
e^{-w(x^{1}+x^{1\prime })}}{w}\frac{\beta w+1}{\beta w-1},  \label{W2}
\end{eqnarray}%
where $w=\sqrt{x^{2}+k^{2}+\gamma ^{2}}$.

The expression for the Wightman function is further simplified for special
cases of Dirichlet and Neumann boundary conditions. The corresponding
integral over $k$ is expressed in terms of the MacDonald function. Next, we
integrate over $x$, that again gives the MacDonald function. And finally,
after the integration over $\gamma $, we come to the expression
\begin{equation}
W(x,x^{\prime })=W_{0}(x,x^{\prime })\mp \frac{\alpha ^{1-D}f_{\nu }(u_{+})}{%
2^{D/2+\nu +1}\pi ^{D/2}},  \label{WDN}
\end{equation}%
where upper/lower signs correspond to Dirichlet/Neumann boundary conditions
and%
\begin{equation}
u_{+}=1+[(\Delta z)^{2}+(x^{1}+x^{1\prime })^{2}+|\Delta \mathbf{x}%
|^{2}-(\Delta t)^{2}]/(2zz^{\prime }).  \label{up}
\end{equation}%
The quantity $u_{+}$ is expressed in terms of the geodesic distance between
the points $(t,x^{1},\mathbf{x},z)$ and $(t^{\prime },-x^{1\prime },\mathbf{x%
}^{\prime },z^{\prime })$. The latter is the image point of $(t^{\prime
},x^{1\prime },\mathbf{x}^{\prime },z^{\prime })$ with respect to the brane.

For points away from the brane the local geometry is the same as that for
the AdS spacetime in the absence of the brane. As a consequence of this, the
divergences in the VEVs of the blinear combinations of the field operator
(field squared, energy-momentum tensor) in the coincidence limit come from
the boundary-free part of the Wightman function. Hence, with the
decompositions (\ref{W2}) and (\ref{WDN}), the renormalization of those VEVs
is reduced to the ones in the boundary-free geometry.

\subsection{Bulk-to-boundary propagator}

By using the mode functions given above we can also evaluate the
bulk-to-boundary propagator which is among the central objects in the
AdS/CFT correspondence. The latter is usually discussed in Euclidean
signature. In terms of the coordinate $z$, the corresponding line element is
written as $ds^{2}=(\alpha /z)^{2}[(dx^{1})^{2}+(d\mathbf{X)}^{2}+dz^{2}]$,
where $\mathbf{X}=(X^{0},\mathbf{x})$. The solutions of the field equation,
which obey the boundary condition (\ref{boundcond}) and do not diverge in
the limit $z\rightarrow \infty $, have the form%
\begin{equation}
\varphi _{E\sigma }(x)=C_{E}z^{D/2}K_{\nu }(\gamma z)\cos [\lambda
x^{1}+\alpha _{0}(\lambda )]e^{i\mathbf{KX}},  \label{phiEsig}
\end{equation}%
where $K_{\nu }(x)$ is the Macdonald function and $\gamma =\sqrt{%
K^{2}+\lambda ^{2}}$. Now, the general solution of the field equation is
presented as%
\begin{equation}
\varphi (\mathbf{X}_{D},z)=\frac{z^{D/2}}{2^{\nu -2}\Gamma (\nu )}\int
d^{D-1}\mathbf{K}d\lambda \gamma ^{\nu }\varphi _{(0)}(\lambda ,\mathbf{K}%
)K_{\nu }(\gamma z)\cos [\lambda x^{1}+\alpha _{0}(\lambda )]e^{i\mathbf{KX}%
},  \label{phiexp}
\end{equation}%
where $\mathbf{X}_{D}=(x^{1},\mathbf{X})$. By taking into account that for $%
\beta <0$ the functions $\cos [\lambda x^{1}+\alpha _{0}(\lambda )]$ form a
complete set we can write
\begin{equation}
\varphi _{(0)}(\lambda ,\mathbf{K})=\frac{1}{2^{D-1}\pi ^{D}}\int d^{D-1}%
\mathbf{X}^{\prime }\,\int_{0}^{\infty }dx^{1\prime }\varphi
_{(0)}(x^{1\prime },\mathbf{X}^{\prime })\cos [\lambda x^{1\prime }+\alpha
_{0}(\lambda )]e^{-i\mathbf{KX}^{\prime }}.  \label{phi0exp}
\end{equation}%
Substituting this into (\ref{phiexp}) one finds the following relation%
\begin{equation}
\varphi (\mathbf{X}_{D},z)=\int d^{D-1}\mathbf{X}^{\prime
}\,\int_{0}^{\infty }dx^{1\prime }G(\mathbf{X}_{D};\mathbf{X}_{D}^{\prime
},z)\varphi _{(0)}(x^{1\prime },\mathbf{X}^{\prime }),  \label{phiexp2}
\end{equation}%
with the bulk-to-boundary propagator%
\begin{eqnarray}
G(\mathbf{X}_{D};\mathbf{X}_{D}^{\prime },z) &=&\frac{2^{3-\nu }z^{D/2}}{%
(2\pi )^{D}\Gamma (\nu )}\int d^{D-1}\mathbf{K}\int_{0}^{\infty }d\lambda
\gamma ^{\nu }K_{\nu }(\gamma z)  \notag \\
&&\times \cos [\lambda x^{1}+\alpha _{0}(\lambda )]\cos [\lambda x^{1\prime
}+\alpha _{0}(\lambda )]e^{i\mathbf{K}\Delta \mathbf{X}},  \label{Gprop}
\end{eqnarray}%
and $\Delta \mathbf{X}=\mathbf{X}-\mathbf{X}^{\prime }$. For small $z$, to
the leading order, from (\ref{phiexp2}) one gets $\varphi (\mathbf{X}%
_{D},z)\approx z^{D/2-\nu }\varphi _{(0)}(x^{1},\mathbf{X})$. In the AdS/CFT
correspondence, $\varphi _{(0)}(x^{1},\mathbf{X})\equiv \varphi _{(0)}(%
\mathbf{X}_{D})$ is interpreted as the source for a dual scalar operator.
Note that the coefficient in (\ref{phiexp}) is chosen so that the expression
$z^{D/2-\nu }\varphi _{(0)}(\mathbf{X}_{D})$ for the leading term is
obtained.

The propagator (\ref{Gprop}) can be presented in the form%
\begin{eqnarray}
G(\mathbf{X}_{D};\mathbf{X}_{D}^{\prime },z) &=&G_{0}(\mathbf{X}_{D};\mathbf{%
X}_{D}^{\prime },z)+\frac{2^{1-\nu }z^{D/2}}{(2\pi )^{D}\Gamma (\nu )}\int
d^{D-1}\mathbf{K}  \notag \\
&&\times \int_{-\infty }^{\infty }d\lambda \,\gamma ^{\nu }K_{\nu }(\gamma z)%
\frac{i\beta \lambda -1}{i\beta \lambda +1}e^{i\lambda (x^{1}+x^{1\prime
})}e^{i\mathbf{K}\Delta \mathbf{X}},  \label{Gprop2}
\end{eqnarray}%
where
\begin{eqnarray}
G_{0}(\mathbf{X}_{D};\mathbf{X}_{D}^{\prime },z) &=&\frac{2^{1-\nu }z^{D/2}}{%
(2\pi )^{D}\Gamma (\nu )}\int d^{D}\mathbf{K}_{D}\,|\mathbf{K}_{D}|^{\nu
}K_{\nu }(|\mathbf{K}_{D}|z)e^{i\mathbf{K}_{D}\Delta \mathbf{X}_{D}}  \notag
\\
&=&\frac{\pi ^{-D/2}\Gamma (\nu +D/2)z^{\nu +D/2}}{\Gamma (\nu )\left(
|\Delta \mathbf{X}_{D}|^{2}+z^{2}\right) ^{\nu +D/2}},  \label{G0prop}
\end{eqnarray}%
with $\Delta \mathbf{X}_{D}=\mathbf{X}_{D}-\mathbf{X}_{D}^{\prime }$, is the
bulk-to-boundary propagator in the geometry without the brane \cite{Free99}.
The second term in the right-hand side of (\ref{Gprop2}) is induced by the
brane.

For Dirichlet and Neumann boundary conditions the brane-induced contribution
in (\ref{Gprop2}) is evaluated as $\mp G_{0}(\mathbf{X}_{D};\mathbf{X}%
_{D}^{(-)\prime },z)$, where $\mathbf{X}_{D}^{(-)\prime }=(-x^{1\prime },%
\mathbf{X}^{\prime })$. This result could be directly obtained by the image
method. The bulk-to-boundary propagator in these relatively simple special
cases has been discussed in \cite{Alis11}.

\section{Mean field squared}

\label{sec:phi2}

The VEV of the field squared is obtained from the Wightman function in the
coincidence limit, $x^{\prime }\rightarrow x$, and is decomposed as%
\begin{equation}
\langle \varphi ^{2}\rangle =\langle \varphi ^{2}\rangle _{0}+\langle
\varphi ^{2}\rangle _{b}.  \label{phi2dec}
\end{equation}%
Here $\langle \varphi ^{2}\rangle _{0}$ is the renormalized VEV in the
boundary-free AdS spacetime and $\langle \varphi ^{2}\rangle _{b}$ is the
contribution induced the brane. By using (\ref{W2}), for the brane-induced
contribution one has the expression
\begin{eqnarray}
\langle \varphi ^{2}\rangle _{b} &=&\frac{4\alpha ^{1-D}z^{D}}{\left( 4\pi
\right) ^{D/2}\Gamma (D/2-1)}\int_{0}^{\infty }dk\,k^{D-3}\int_{0}^{\infty
}d\gamma \,\gamma J_{\nu }^{2}(\gamma z)\,  \notag \\
&&\times \int_{0}^{\infty }dx\,\frac{e^{-2wx^{1}}}{w}\frac{\beta w+1}{\beta
w-1},  \label{phi2}
\end{eqnarray}%
with $w=\sqrt{x^{2}+k^{2}+\gamma ^{2}}$. As a consequence of the maximal
symmetry of AdS spacetime, the boundary-free part does not depend on the
spacetime point. This VEV has been investigated in the literature \cite%
{Burg85}-\cite{Cald99}. Here, we shall be mainly concerned with the
boundary-induced part, given by (\ref{phi2}).

For the further transformation of the brane-induced contribution we
introduce polar coordinates $(r,\theta )$ in the plane $(k,x)$. The
integration over the angle $\theta $ is done explicitly. Then, we introduce
polar coordinates $(w,\theta ^{\prime })$ in the plane $(r,\gamma )$.
Introducing a new integration variable $\tau =\sin \theta ^{\prime }$, we get%
\begin{equation}
\langle \varphi ^{2}\rangle _{b}=\frac{\left( 4\pi \right) ^{(1-D)/2}z^{D}}{%
\Gamma ((D-1)/2)\alpha ^{D-1}}\int_{0}^{\infty }dw\,w^{D-1}e^{-2wx^{1}}\frac{%
\beta w+1}{\beta w-1}\int_{0}^{1}d\tau \,\tau (1-\tau ^{2})^{(D-3)/2}J_{\nu
}^{2}(zw\tau ).  \label{phi22}
\end{equation}%
For the integral over $\tau $ one has \cite{Prud2}%
\begin{equation}
\int_{0}^{1}d\tau \,\tau (1-\tau ^{2})^{(D-3)/2}J_{\nu }^{2}(u\tau )=\frac{%
\Gamma ((D-1)/2)}{2^{2\nu +1}}u^{2\nu }F_{\nu }^{D/2}(u),  \label{Int1}
\end{equation}%
with the notation%
\begin{equation}
F_{\nu }^{\mu }(u)=\frac{\,_{1}F_{2}\left( \nu +1/2;\nu +\mu +1/2,2\nu
+1;-u^{2}\right) }{\Gamma (\nu +\mu +1/2)\Gamma (\nu +1)},  \label{Fnu}
\end{equation}%
where $_{1}F_{2}$ is the hypergeometric function. So, for the brane-induced
contribution one gets%
\begin{equation}
\langle \varphi ^{2}\rangle _{b}=\frac{2^{-2\nu -1}\alpha ^{1-D}}{\left(
4\pi \right) ^{(D-1)/2}}\int_{0}^{\infty }dx\,x^{D+2\nu
-1}e^{-2xx^{1}/z}F_{\nu }^{D/2}(x)\frac{x\beta /z+1}{x\beta /z-1}.
\label{phi23}
\end{equation}%
As is seen, the VEV depends on $x^{1}$, $\beta $, $z$, in the form of the
dimensionless ratios $x^{1}/z$ and $\beta /z$. This property is a
consequence of the maximal symmetry of AdS spacetime. Note that the ratio $%
x^{1}/z$ is the proper distance from the brane, $\alpha x^{1}/z$, measured
in units of the AdS curvature radius $\alpha $.

In the case of Dirichlet and Neumann boundary conditions we get%
\begin{equation}
\langle \varphi ^{2}\rangle _{b}=\mp \frac{\alpha ^{1-D}f_{\nu }(u)}{%
2^{D/2+\nu +1}\pi ^{D/2}},  \label{phi2DN}
\end{equation}%
where the notation%
\begin{equation}
u=1+2(x^{1}/z)^{2}  \label{uD}
\end{equation}%
is used. For these boundary conditions the VEV of the field squared is a
function of the proper distance from the brane alone. As is seen from (\ref%
{phi23}), for a fixed value of the proper distance from the brane, $\alpha
x^{1}/z$, near the AdS boundary, $z\rightarrow 0$, the Neumann boundary
condition is an "attractor" in the general class of boundary conditions
specified by the parameter $\beta $, whereas Dirichlet boundary condition is
an "attractor" near the horizon, corresponding to $z\rightarrow \infty $.

The VEV of the field squared for a plate in Minkowski spacetime is obtained
from (\ref{phi23}) in the limit $\alpha \rightarrow \infty $ for a fixed
value of the coordinate $y$. In this case, in the leading order, one has $%
\nu \approx m\alpha $ and$\;z\approx \alpha $. Introducing in (\ref{phi23})
a new integration variable $w=x/\alpha $, we see that in the limit under
consideration both the order and the argument of the function $F_{\nu
}^{D/2}(\alpha w)\approx F_{\nu }^{D/2}(\nu w/m)$ are large. The
corresponding uniform asymptotic expansion can be obtained by making use of
the relation (\ref{Int1}) and the expansion for the Bessel function. In this
way it can be seen that for $w<m$ the function $F_{\nu }^{D/2}(\nu w/m)$ is
exponentially suppressed for large $\nu $ and the dominant contribution to
the integral for $\langle \varphi ^{2}\rangle _{b}$ comes from the region $%
w>m$. In this region, for $\nu \gg 1$, to the leading order one gets%
\begin{equation}
F_{\nu }^{D/2}(\nu w/m)\approx \frac{2^{2\nu }[1-\left( m/w\right)
^{2}]^{D/2-1}}{\sqrt{\pi }\Gamma (D/2)(\nu w/m)^{2\nu +1}}.
\label{FnuLarge2}
\end{equation}%
Substituting this into the expression for the field squared, we find $%
\langle \varphi ^{2}\rangle _{b}\approx \langle \varphi ^{2}\rangle
_{b}^{(M)}$ with%
\begin{equation}
\langle \varphi ^{2}\rangle _{b}^{(M)}=\frac{\left( 4\pi \right) ^{-D/2}}{%
\Gamma (D/2)}\int_{m}^{\infty }dx\,e^{-2xx^{1}}\left( x^{2}-m^{2}\right)
^{D/2-1}\frac{x\beta +1}{x\beta -1},  \label{Phi2Mink}
\end{equation}%
being the corresponding VEV for a plate in Minkowski spacetime (for the VEVs
in the geometry of a single and two parallel Robin plates in Minkowski
spacetime see \cite{Rome02}).

The general expression of the field squared, given by (\ref{phi23}), is
simplified in the asymptotic regions. At small proper distances from the
brane, compared with the AdS curvature radius, one has $x^{1}/z\ll 1$ and
the dominant contribution in (\ref{phi23}) comes from large values of $x$.
For these values one has the asymptotic expression
\begin{equation}
F_{\nu }^{\mu }(x)\approx \frac{2^{2\nu }x^{-2\nu -1}}{\sqrt{\pi }\Gamma
(\mu )},\;x\gg 1,  \label{FnuLarge}
\end{equation}%
and from (\ref{phi23}), in the leading order, we find%
\begin{equation}
\langle \varphi ^{2}\rangle _{b}\approx \left( z/\alpha \right)
^{D-1}\langle \varphi ^{2}\rangle _{b}^{(M)}|_{m=0}.  \label{phi2bnear}
\end{equation}%
If, in addition, $x^{1}/|\beta |\ll 1$, one gets%
\begin{equation}
\langle \varphi ^{2}\rangle _{b}\approx \frac{\Gamma ((D-1)/2)}{\left( 4\pi
\right) ^{(D+1)/2}}\left( \frac{z}{\alpha x^{1}}\right) ^{D-1}.
\label{phi2bnear2}
\end{equation}%
For Dirichlet boundary condition, $\beta =0$, the leading term in the
asymptotic expansion near the brane is given by (\ref{phi2bnear2}) with the
opposite sign. Note that, for a fixed value $x^{1}$, the expression in the
right-hand side of (\ref{phi2bnear2}) provides the leading term near the AdS
horizon, $z\rightarrow \infty $. As is seen, for $x^{1}\neq 0$, the
brane-induced VEV diverges on the horizon as $z^{D-1}$.

At large proper distances from the brane compared with the AdS curvature
radius, $x^{1}/z\gg 1$, the dominant contribution in (\ref{phi23}) comes
from small values of $x$. By taking into account that%
\begin{equation}
F_{\nu }^{D/2}(x)\approx \frac{\,1}{\Gamma ((D+1)/2+\nu )\Gamma (1+\nu )}%
,\;x\ll 1,  \label{FnuSmall}
\end{equation}%
and assuming that $|\beta |/x^{1}\ll 1$ to the leading order one gets%
\begin{equation}
\langle \varphi ^{2}\rangle _{b}\approx -\frac{\alpha ^{1-D}\Gamma (D/2+\nu )%
}{2\pi ^{D/2}\Gamma (1+\nu )(2x^{1}/z)^{D+2\nu }}.  \label{phi2Large}
\end{equation}%
For Neumann boundary condition, $\beta =\infty $, the leading term coincides
with (\ref{phi2Large}) with the opposite sign. Note that the decay of the
boundary-induced contribution at large distances from the brane, as a
function of the proper distance $\alpha x^{1}/z$, is power-law for both
massless and massive fields. This is in clear contrast with the case of the
problem in Minkowski bulk (for a similar feature for a Robin boundary in de
Sitter spacetime see \cite{Saha09}). In the latter geometry the
boundary-induced VEV (see (\ref{Phi2Mink})) decays as $1/(x^{1})^{D-1}$ for
a massless field and is exponentially suppressed (as $e^{-2mx^{1}}$) for a
massive field. From (\ref{phi2Large}) it follows that, for a given $x^{1}$,
the brane-induced contribution in the VEV of the field squared vanishes on
the AdS boundary as $z^{D+2\nu }$. Note that the quantity $\alpha x^{1}/z$
is the proper distance from the brane measured by an observer with a fixed
value of the coordinate $x^{1}$. This observer is at rest with respect to
the brane. The geodesic distance $\sigma (x,x^{\prime })$ between the points
$x=(t,0,\mathbf{x},z)$ and $x^{\prime }=(t,x^{1},\mathbf{x},z)$ is given by
the relation $\cosh (\sigma (x,x^{\prime })/\alpha )=1+(x^{1}/z)^{2}/2$. At
large distances from the brane one gets $(x^{1}/z)^{2}=e^{\sigma
(x,x^{\prime })/\alpha }$.

\begin{figure}[tbph]
\begin{center}
\begin{tabular}{cc}
\epsfig{figure=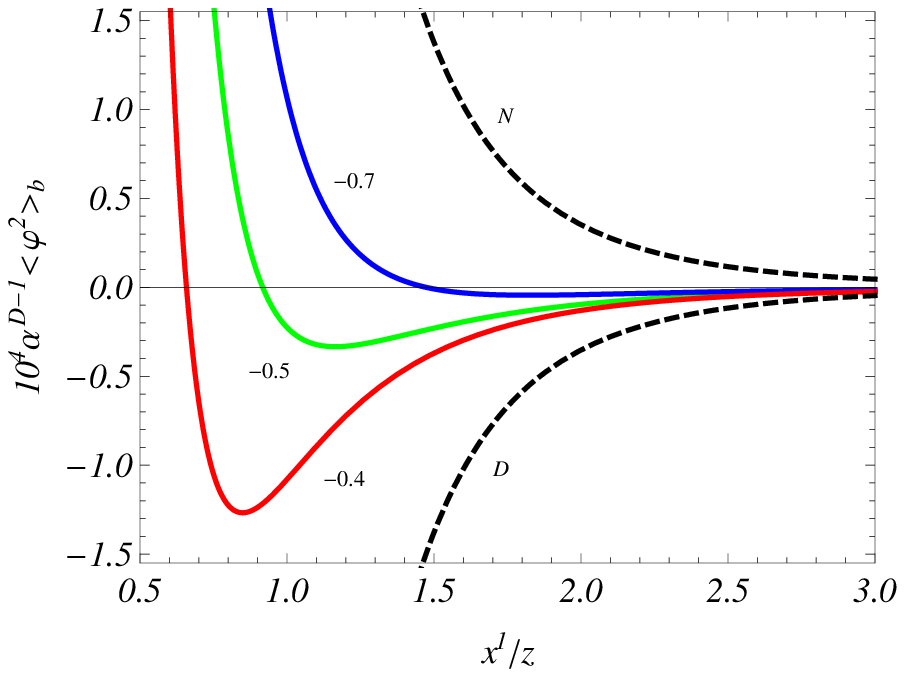,width=7.cm,height=5.5cm} & \quad %
\epsfig{figure=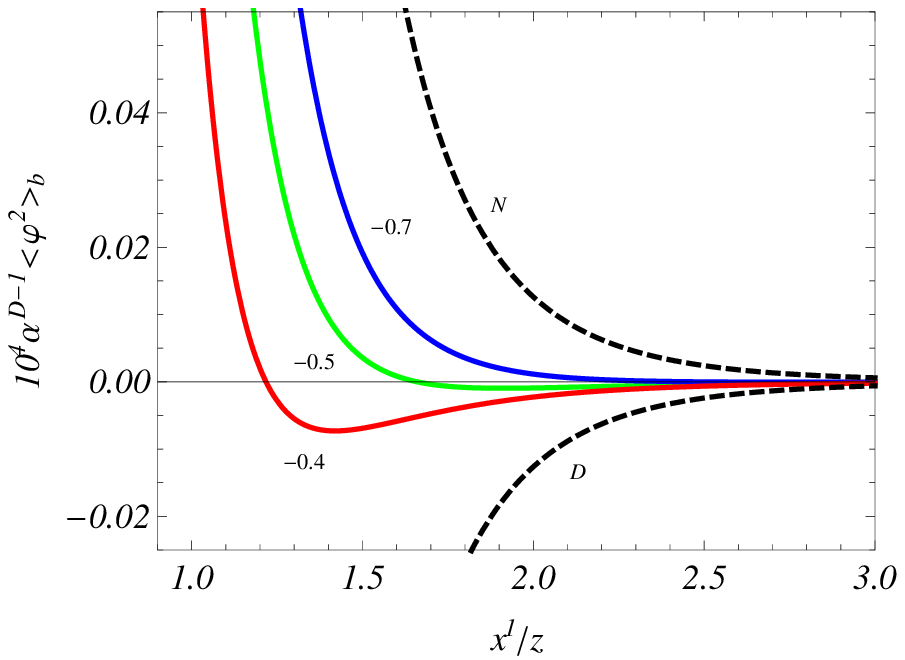,width=7.cm,height=5.5cm}%
\end{tabular}%
\end{center}
\caption{The boundary-induced part in the VEV of the field squared versus
the proper distance from the brane for $D=4$ conformally (left panel) and
minimally (right panel) coupled scalar fields with $m\protect\alpha =0.5$.
The numbers near the solid curves correspond to the values of the ratio $%
\protect\beta /z$ and the dashed curves are for Dirichlet and Neumann
boundary conditions.}
\label{fig1}
\end{figure}

In figure \ref{fig1} we have plotted the brane-induced contribution in the
VEV of the field squared for $D=4$ conformally (left panel) and minimally
(right panel) coupled scalar fields, as a function of the proper distance
from the brane (measured in units of the AdS curvature radius). The numbers
near the curves correspond to the values of the ratio $\beta /z$. The dashed
lines correspond to Dirichlet and Neumann boundary conditions. The graphs
are plotted for $m\alpha =0.5$. A feature obtained from the asymptotic
analysis above is seen: Neumann boundary condition is an "attractor" in a
general class of Robin conditions for points near the brane, whereas
Dirichlet boundary condition is an "attractor" at large distances.

\section{Vacuum energy density and stresses}

\label{sec:EMT}

Having the Wightman function and the mean field squared, the VEV of the
energy-momentum tensor is evaluated by making use of the formula
\begin{equation}
\langle T_{ik}\rangle =\lim_{x^{\prime }\rightarrow x}\partial _{i}\partial
_{k}^{\prime }W(x,x^{\prime })+\left[ \left( \xi -1/4\right) g_{ik}\nabla
_{l}\nabla ^{l}-\zeta \nabla _{i}\nabla _{k}-\zeta R_{ik}\right] \langle
\varphi ^{2}\rangle ,  \label{vevEMT1pl}
\end{equation}%
where $R_{ik}=-Dg_{ik}/\alpha ^{2}$ is the Ricci tensor for AdS spacetime.
In defining the right-hand side of this formula we have used the expression
for the energy-momentum tensor for a scalar field which differs from the
standard one (given, for example, in \cite{Birr82}) by a term that vanishes
on the solutions of the field equation (see \cite{Saha04}). Similar to the
VEV of the field squared, the vacuum energy-momentum tensor is decomposed
into boundary-free and boundary-induced parts:%
\begin{equation}
\langle T_{ik}\rangle =\langle T_{ik}\rangle _{0}+\langle T_{ik}\rangle _{b}.
\label{Tikdec}
\end{equation}%
As a consequence of the maximal symmetry of boundary-free AdS spacetime and
of the vacuum state under consideration, one has $\langle T_{ik}\rangle _{0}=%
\mathrm{const}\cdot g_{ik}$. Hence, the corresponding vacuum energy-momentum
tensor is completely determined by its trace. The boundary-free
energy-momentum tensor $\langle T_{ik}\rangle _{0}$ is well investigated in
the literature (see, for instance, \cite{Camp92}) and in what follows we
shall be concerned with the brane-induced contribution, $\langle
T_{ik}\rangle _{b}$.

For the covariant d'Alembertian acted on the brane-induced part of the field
squared we find%
\begin{equation}
\nabla _{l}\nabla ^{l}\langle \varphi ^{2}\rangle _{b}=-\frac{2^{-2\nu
-1}\alpha ^{-D-1}}{\left( 4\pi \right) ^{(D-1)/2}}\int_{0}^{\infty
}dx\,xe^{-2xx^{1}/z}\frac{\beta x/z+1}{\beta x/z-1}\hat{B}(x)x^{D+2\nu
}F_{\nu }^{D/2}(x),  \label{Dalamb}
\end{equation}%
with the differential operator%
\begin{equation}
\hat{B}(x)=\partial _{x}^{2}-\frac{D-1}{x}\partial _{x}+4.  \label{Boper}
\end{equation}

By making use of the expressions for the boundary-induced contributions in
the Wightman function and in the VEV of the field squared, Eq. (\ref{phi23}%
), from (\ref{vevEMT1pl}) for the diagonal components in the region $x^{1}>0$
one gets (no summation over $i=0,1,\ldots ,D$)%
\begin{eqnarray}
\langle T_{i}^{i}\rangle _{b} &=&-\frac{\left( 4\pi \right) ^{(1-D)/2}}{%
2^{2\nu +1}\alpha ^{D+1}}\int_{0}^{\infty }dx\,xe^{-2xx^{1}/z}\frac{\beta
x/z+1}{\beta x/z-1}  \notag \\
&&\times \left[ A_{i}x^{D+2\nu }F_{\nu }^{D/2+1}(x)+\hat{B}_{i}(x)x^{D+2\nu
}F_{\nu }^{D/2}(x)\right] ,  \label{Tii}
\end{eqnarray}%
with the notations%
\begin{eqnarray}
A_{i} &=&1/2,\;i=0,2,\ldots ,D-1,  \notag \\
A_{1} &=&0,\;A_{D}=\frac{1-D}{2},  \label{Ai}
\end{eqnarray}%
and%
\begin{eqnarray}
\hat{B}_{i}(x) &=&\frac{\xi _{1}}{4}\hat{B}(x)+\frac{\xi }{x}\partial _{x}-%
\frac{\xi D}{x^{2}},\;i=0,2,\ldots ,D-1,  \notag \\
\hat{B}_{1}(x) &=&\frac{\xi _{1}}{4}\hat{B}(x)+\frac{\xi }{x}\partial _{x}-%
\frac{\xi D}{x^{2}}-\xi _{1},  \label{Bi} \\
\hat{B}_{D}(x) &=&\frac{1}{4}\hat{B}(x)-\xi \frac{D}{x}\partial _{x}+\xi
_{1}+\frac{D^{2}\xi -m^{2}\alpha ^{2}}{x^{2}}.  \notag
\end{eqnarray}%
Here and in what follows, we use the notation $\xi _{1}=4\xi -1$. The
diagonal components are symmetric under the reflection $x^{1}\rightarrow
-x^{1}$ with respect to the brane: they are given by the expression (\ref%
{Tii}) with $x^{1}$ replaced by $|x^{1}|$. The vacuum stresses along the
directions parallel to the brane are equal to the energy density. This
property is a consequence of the invariance with respect to the Lorentz
boosts along those directions.

In addition to the diagonal components, the vacuum energy-momentum tensor
has off-diagonal components $\langle T_{D}^{1}\rangle _{b}=\langle
T_{1}^{D}\rangle _{b}$. For the latter one gets%
\begin{equation}
\langle T_{D}^{1}\rangle _{b}=\frac{\left( 4\pi \right) ^{(1-D)/2}}{2^{2\nu
}\alpha ^{D+1}}\int_{0}^{\infty }dx\,e^{-2xx^{1}/z}\frac{\beta x/z+1}{\beta
x/z-1}\left[ \left( \frac{1}{4}-\xi \right) x\partial _{x}-\xi \right]
x^{D+2\nu }F_{\nu }^{D/2}(x).  \label{T1D}
\end{equation}%
This off-diagonal component changes the sign under the reflection $%
x^{1}\rightarrow -x^{1}$. Similar to the case of the field squared, the mean
energy-momentum tensor depends on the coordinates $x^{1}$, $z$, and on the
parameter $\beta $ in the form of the ratios $x^{1}/z$ and $\beta /z$. The
first of these is the proper distance from the brane measured in units of
the curvature radius $\alpha $. Note that for the derivatives appearing in (%
\ref{Tii}) and (\ref{T1D}) one has the relations%
\begin{eqnarray}
\partial _{x}[x^{D+2\nu }F_{\nu }^{D/2}(x)] &=&x^{D+2\nu -1}[F_{\nu
}^{D/2}(x)+2F_{\nu }^{D/2-1}(x)],  \notag \\
\partial _{x}^{2}[x^{D+2\nu }F_{\nu }^{D/2}(x)] &=&2x^{D+2\nu -2}[3F_{\nu
}^{D/2-1}(x)+2F_{\nu }^{D/2-2}(x)].  \label{DerF}
\end{eqnarray}
Here we have used the formula $\partial _{z}\left[ z^{b}\,_{1}F_{2}\left(
a;b+1,2a;z\right) \right] =bz^{b-1}\,_{1}F_{2}\left( a;b,2a;z\right) $ with $%
a=\nu +1/2$.

By using the expressions given above, we can see that the boundary-induced
contributions obey the trace relation
\begin{equation}
\langle T_{i}^{i}\rangle _{b}=D\left( \xi -\xi _{D}\right) \nabla _{l}\nabla
^{l}\langle \varphi ^{2}\rangle _{b}+m^{2}\langle \varphi ^{2}\rangle _{b}.
\label{TraceRel}
\end{equation}%
In particular, the brane-induced part is traceless for a conformally coupled
massless field. The trace anomalies are contained in the boundary-free part
only. As an additional check for the expressions given above, we can see
that the covariant continuity equation $\nabla _{k}\langle T_{i}^{k}\rangle
_{b}=0$ is obeyed. For the geometry under consideration the latter is
reduced to the following relations%
\begin{eqnarray}
\partial _{1}\langle T_{1}^{1}\rangle _{b}+\partial _{D}\langle
T_{1}^{D}\rangle _{b}-\frac{D+1}{z}\langle T_{1}^{D}\rangle _{b} &=&0,
\notag \\
\partial _{1}\langle T_{D}^{1}\rangle _{b}+\partial _{D}\langle
T_{D}^{D}\rangle _{b}-\frac{D}{z}\langle T_{D}^{D}\rangle _{b}+\frac{1}{z}%
\sum_{k=0}^{D-1}\langle T_{k}^{k}\rangle _{b} &=&0.  \label{ContEq}
\end{eqnarray}

The Minkowskian limit for the VEVs of the energy-momentum tensor is
considered in a way similar to that for the VEV of the field squared. By
using the asymptotic expression (\ref{FnuLarge2}), to the leading order for
the diagonal components we get (nu summation over $i$) $\langle
T_{i}^{i}\rangle _{b}\approx \langle T_{i}^{i}\rangle _{b}^{(M)}$, where for
a plate in Minkowski spacetime one has (see \cite{Rome02})%
\begin{equation}
\langle T_{i}^{i}\rangle _{b}^{(M)}=-\frac{\left( 4\pi \right) ^{-D/2}}{%
\Gamma (D/2)}\int_{m}^{\infty }du\,e^{-2xx^{1}}\frac{\beta u+1}{\beta u-1}%
\left( u^{2}-m^{2}\right) ^{D/2-1}\left[ 4\left( \xi -\xi _{D}\right)
u^{2}-m^{2}/D\right] ,  \label{TiiM}
\end{equation}%
for $i=0,2,\ldots ,D$ and $\langle T_{1}^{1}\rangle _{b}^{(M)}=0$. For the
leading term in the off-diagonal component we find%
\begin{eqnarray}
\langle T_{D}^{1}\rangle _{b} &\approx &-\frac{2\left( 4\pi \right) ^{-D/2}}{%
\Gamma (D/2)\alpha }\int_{0}^{\infty }du\,ue^{-2ux^{1}}\frac{\beta u+1}{%
\beta u-1}\left( u^{2}-m^{2}\right) ^{D/2-2}  \notag \\
&&\times \left[ D\left( \xi -\xi _{D}\right) u^{2}+\left( 1/4-2\xi \right)
m^{2}\right] ,  \label{T1DM}
\end{eqnarray}%
and it vanishes in the Minkowskian limit. Note that for the normal stress
one has $\langle T_{1}^{1}\rangle _{b}=\mathcal{O}(1/\alpha ^{2})$. For a
conformally coupled massless field, the Minkowskian limit of the VEV of the
energy-momentum tensor vanishes for a single plate.

In the case of Dirichlet and Neumann boundary conditions it is convenient to
evaluate the VEV of the energy-momentum tensor by using the formula (\ref%
{vevEMT1pl}) with the Wightman function from (\ref{WDN}). For the
boundary-induced contributions in the diagonal components we find (no
summation over $i$)%
\begin{equation}
\langle T_{i}^{i}\rangle _{b}=\pm \frac{\pi ^{-D/2}\alpha ^{-D-1}}{%
2^{D/2+\nu +1}}[\hat{C}_{i}(u)-D\xi ]f_{\nu }(u),  \label{TiiDN}
\end{equation}%
where, as before, the upper/lower sign corresponds to Dirichlet/Neumann
boundary condition and $u$ is defined in accordance with (\ref{uD}). In (\ref%
{TiiDN}), $\hat{C}_{i}(u)$ are the second order differential operators
defined by the expressions ($i=0,2,\ldots ,D-1$)%
\begin{eqnarray}
\hat{C}_{i}(u) &=&\xi _{1}\left( u^{2}-1\right) \partial _{u}^{2}+\left[
4\xi -2+\left( \frac{D+1}{2}\xi _{1}-\frac{1}{2}\right) \left( u-1\right) %
\right] \partial _{u},  \notag \\
\hat{C}_{1}(u) &=&\xi _{1}\left( u-1\right) ^{2}\partial _{u}^{2}+\left(
\frac{D+1}{2}\xi _{1}-\frac{1}{2}\right) \left( u-1\right) \partial _{u},
\notag \\
\hat{C}_{D}(u) &=&2\xi _{1}\left( u-1\right) \partial _{u}^{2}+\left[ 4\xi
-2+\frac{D}{2}\xi _{1}\left( u-1\right) \right] \partial _{u},  \label{CD}
\end{eqnarray}%
with $\xi _{1}=4\xi -1$. For the off-diagonal component one gets%
\begin{equation}
\langle T_{D}^{1}\rangle _{b}=\pm \frac{2\alpha ^{-D-1}x^{1}/z}{2^{D/2+\nu
+1}\pi ^{D/2}}\left[ \xi _{1}\left( u-1\right) \partial _{u}^{2}+\left( 2\xi
-1\right) \partial _{u}\right] f_{\nu }(u).  \label{T1DD}
\end{equation}%
The second derivatives in (\ref{TiiDN}) and (\ref{T1DD}) can be excluded by
using the differential equation for the function $f_{\nu }(u)$. The latter
is obtained by using the definition (\ref{fnu}) and the equation for the
function $\,{}_{2}F_{1}$. In this way we can see that%
\begin{equation}
\lbrack \left( u^{2}-1\right) \partial _{u}^{2}+\left( D+1\right) u\partial
_{u}+D^{2}/4-\nu ^{2}]f_{\nu }(u)=0.  \label{fnuEq}
\end{equation}%
As an additional check, by making use of (\ref{fnuEq}), it can be shown that
the VEVs (\ref{TiiDN}) and (\ref{T1DD}) obey the trace relation (\ref%
{TraceRel}).

Let us consider the asymptotic behavior of the vacuum energy-momentum tensor
near the brane and at large distances for general case of Robin boundary
condition. For points near the brane, $x^{1}/z\ll 1$, the dominant
contribution to the integral in (\ref{Tii}) comes from large values of $x$.
The corresponding asymptotic of the function $F_{\nu }^{\mu }(x)$ was given
by (\ref{FnuLarge}). Assuming that $x^{1}/|\beta |\ll 1$, in the leading
order we find (no summation over $i=0,2,\ldots ,D$)%
\begin{eqnarray}
\langle T_{i}^{i}\rangle _{b} &\approx &\frac{2D\left( \xi _{D}-\xi \right)
\Gamma ((D+1)/2)}{\pi ^{(D+1)/2}\left( 2\alpha x^{1}/z\right) ^{D+1}},
\notag \\
\langle T_{1}^{1}\rangle _{b} &\approx &-\frac{D\left( \xi _{D}-\xi \right)
\Gamma ((D-1)/2)}{4\pi ^{(D+1)/2}\left( 2\alpha x^{1}/z\right) ^{D-1}},
\label{TiiNear}
\end{eqnarray}%
for the diagonal components and
\begin{equation}
\langle T_{D}^{1}\rangle _{b}\approx \frac{D\left( \xi _{D}-\xi \right)
\Gamma ((D+1)/2)}{\pi ^{(D+1)/2}\alpha (2\alpha x^{1}/z)^{D}},
\label{T1DNear}
\end{equation}%
for the off-diagonal component. For Dirichlet boundary condition ($\beta =0$%
) the asymptotic expressions are given by (\ref{TiiNear}) and (\ref{T1DNear}%
) with the opposite signs. For fixed $x^{1}$, the expressions in (\ref%
{TiiNear}) and (\ref{T1DNear}) give the leading terms near the AdS horizon.
In particular, from (\ref{TiiNear}) it follows that the energy density
diverges on the horizon as $z^{D+1}$. Note that in the evaluation of the
total energy induced by the brane, $E_{b}=\int d^{D}x\,\sqrt{|g|}\langle
T_{0}^{0}\rangle _{b}$, an additional factor $1/z^{D+1}$ comes from the
volume element.

At large distances from the brane, $x^{1}/z\gg 1$, the main contribution to
the integrals in (\ref{Tii}) and (\ref{T1D}) comes from the region near the
lower limit of the integration. Assuming that $|\beta |/x^{1}\ll 1$, for the
diagonal components we get (no summation over $i$)%
\begin{equation}
\langle T_{i}^{i}\rangle _{b}\approx \frac{\alpha ^{-D-1}B_{i}\Gamma
(D/2+\nu )}{\pi ^{D/2}\Gamma (\nu )(2x^{1}/z)^{D+2\nu }},  \label{TiiLarge}
\end{equation}%
where $B_{i}=\left( \xi -1/4\right) \left( D+2\nu \right) +\xi $ for $%
i=0,\ldots ,D-1$, and $B_{D}=-DB_{0}/(2\nu )$. For the off-diagonal
component one finds%
\begin{equation}
\langle T_{D}^{1}\rangle _{b}\approx \frac{2\alpha ^{-D-1}B_{0}\Gamma
(D/2+\nu +1)}{\pi ^{D/2}\Gamma (1+\nu )\left( 2x^{1}/z\right) ^{D+2\nu +1}}.
\label{T1DLarge}
\end{equation}%
As is seen, at large distances the off-diagonal component is suppressed by
an additional factor $x^{1}/z$. For Neumann boundary condition, $\beta =0$,
the asymptotics at large distances are given by the expressions (\ref%
{TiiLarge}) and (\ref{T1DLarge}) with the opposite signs. As in the case of
the field square, at large distances one has a power-law decay instead of
exponential one for the problem with a massive field in Minkowski bulk. From
(\ref{TiiLarge}) and (\ref{T1DLarge}) it follows that, for fixed $x^{1}$,
the diagonal components vanish on the AdS boundary as $z^{D+2\nu }$. The
integrand in the expression for the total energy induced by the brane, $%
E_{b} $, near the horizon behaves like $z^{2\nu -1}$ and the integral over $%
z $ converges at $z=0$ for $\nu >0$.

\begin{figure}[tbph]
\begin{center}
\begin{tabular}{cc}
\epsfig{figure=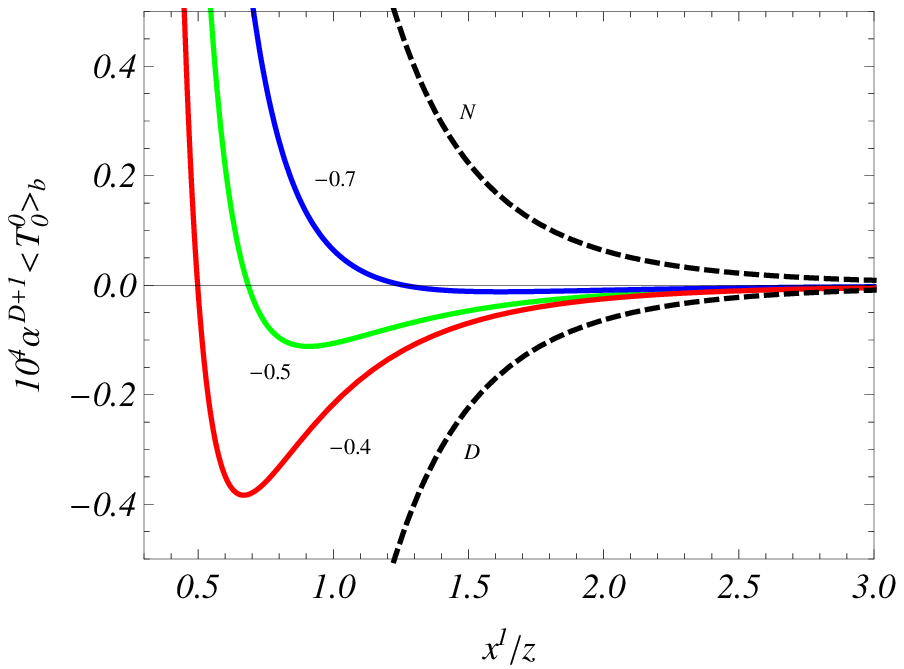,width=7.cm,height=5.5cm} & \quad %
\epsfig{figure=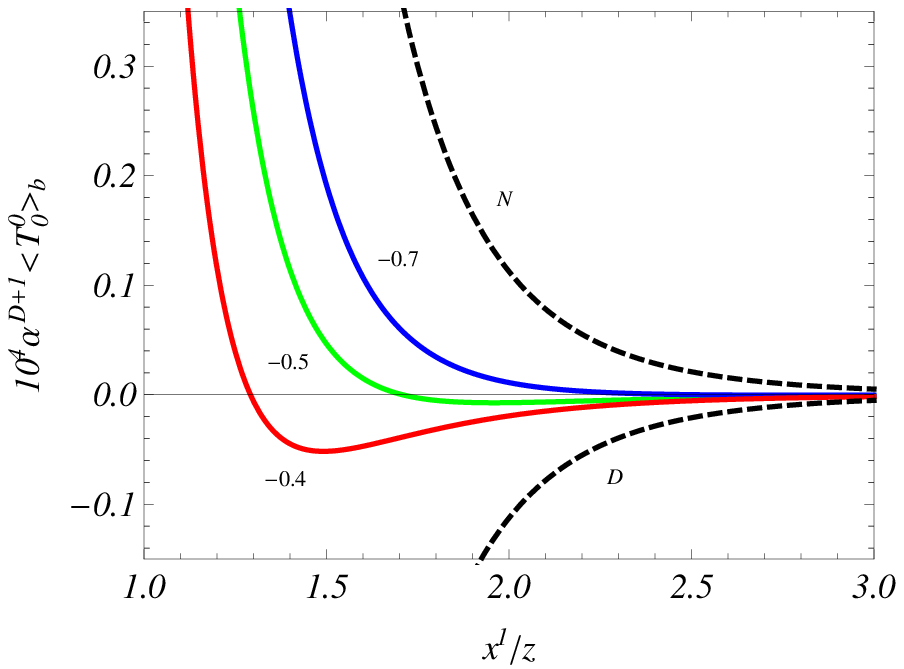,width=7.cm,height=5.5cm}%
\end{tabular}%
\end{center}
\par
.
\caption{The vacuum energy density induced by a brane as a function of the
proper distance measured in units of the AdS curvature radius. The graphs
are plotted for $D=4$ conformally (left panel) and minimally (right panel)
coupled scalar fields with $m\protect\alpha =0.5$. Dashed curves correspond
to Dirichlet and Neumann boundary conditions and the numbers near the solid
curves are the values of the ratio $\protect\beta /z$.}
\label{fig2}
\end{figure}

The figure \ref{fig2} displays the boundary-induced part in the VEV of the
energy density for the cases of $D=4$ conformally (left panel) and minimally
(right panel) coupled scalar fields as a function of the ration $x^{1}/z$
(proper distance from the brane measured in units of the AdS curvature
radius). The dashed lines correspond to Dirichlet and Neumann boundary
conditions. The numbers near the solid curves correspond to the values of
the ratio $\beta /z$. The graphs are plotted for $m\alpha =0.5$.

\section{Conclusion}

\label{sec:Conc}

In the present paper we have investigated quantum effects induced by a flat
brane for a scalar field in background of AdS spacetime. The brane is
perpendicular to the AdS boundary and the field operator obeys Robin
boundary condition on it. We consider a free field theory in AdS spacetime
and all the information on the vacuum state is contained in the two-point
functions. As such a function, the positive-frequency Wightman function is
chosen, which also determines the response of the Unruh-DeWitt-type particle
detectors. We have provided an expression for the Wightman function in which
the contribution induced by the brane is explicitly separated from the pure
AdS one and is given by the second term in the right-hand side of (\ref{W2}%
). This allows to reduce the renormalization procedure for the local VEVs,
at points away from the brane, to the one in AdS spacetime in the absence of
the brane. The latter problem is well discussed in the literature. The
expression for the Wightman function is further simplified in special cases
of Dirichlet and Neumann boundary conditions and is given by (\ref{WDN}).
For a fixed value of the proper distance from the brane, near the AdS
boundary, the Neumann boundary condition is an "attractor" in the general
class of Robin boundary conditions, whereas Dirichlet boundary condition is
an "attractor" near the horizon. We have also evaluated the bulk-to-boundary 
propagator which plays an important role in the discussions of the AdS/CFT
correspondence. Similar to the case of the Wightman function, the corresponding
expression is decomposed into the boundary-free and brane-induced contributions.

As an important characteristic of the quantum vacuum, in section \ref%
{sec:phi2} we have studied the mean field squared. The brane-induced
contribution in this VEV is presented in the form (\ref{phi23}) where the
function $F_{\nu }^{\mu }(x)$ is defined by (\ref{Fnu}). This contribution
depends on the coordinates $x^{1}$, $z$ and on the paramater $\beta $ in the
Robin boundary condition in the form of the ratios $x^{1}/z$ and $\beta /z$.
This property is a consequence of the maximal symmetry of AdS spacetime. For
Dirichlet and Neumann boundary conditions the integral in (\ref{phi23}) is
expressed in terms of the hypergeometric function and the corresponding
formula simplifies to (\ref{phi2DN}). As an additional check of the results
derived, we have shown that in the limit $\alpha \rightarrow \infty $ the
corresponding expression for a Robin plate in Minkowski spacetime is
obtained. The boundary-induced VEV diverges on the brane with the leading
term given by (\ref{phi2bnear2}) for non-Dirichlet boundary conditions. For
Dirichlet boundary condition the leading term has the opposite sign. For
points near the brane, the influence of the gravitational field on the VEV
is small and the leading term coincides with that in Minkowski spacetime.
The influence of gravity is crucial at the proper distances from the brane
larger than the AdS curvature radius. In this limit, for non-Neumann
boundary conditions the leading term in the corresponding asymptotic
expansion has the form (\ref{phi2Large}). For Neumann boundary condition the
same expression is obtained with the opposite sign. For AdS bulk the decay
of the boundary-induced contribution at large distances from the brane is
power-law for both massless and massive fields. This is in clear contrast
with the case of the problem in Minkowski spacetime, where the
boundary-induced VEV for a massive field decays exponentially. For a given $%
x^{1}$, the brane-induced contribution in the VEV of the field squared
vanishes on the AdS boundary as $z^{D+2\nu }$ and diverges on the horizon
like $z^{D-1}$.

Another important quantity, characterizing the vacuum fluctuations in the
presence of the brane, is the VEV of the energy-momentum tensor. The
boundary-induced contributions in the diagonal components are given by (\ref%
{Tii}). The vacuum stresses along the directions parallel to the brane are
equal to the energy density. In addition to the diagonal components the
vacuum energy-momentum tensor has an off-diagonal component defined by the
expression (\ref{T1D}). The formulas for the components of the vacuum
energy-momentum tensor are further simplified for the cases of Dirichlet and
Neumann boundary conditions (see (\ref{TiiDN}) and (\ref{T1DD})). We have
explicitly checked that the brane-induced parts obey the trace relation (\ref%
{TraceRel}) and the covariant conservation equation. The latter is reduced
to the relations (\ref{ContEq}). In the limit of large values for the AdS
curvature radius, to the leading order, for the energy density and parallel
stresses we obtain the corresponding result in Minkowski bulk. In this
limit, the off diagonal component and the normal stress behave like $\langle
T_{D}^{1}\rangle _{b}\propto 1/\alpha $ and $\langle T_{1}^{1}\rangle
_{b}\propto 1/\alpha ^{2}$. For proper distances from the brane smaller than
the AdS curvature radius, with an additional assumption that $x^{1}/|\beta
|\ll 1$, the leading terms in the asymptotic expansion over the distance are
given by (\ref{TiiNear}), (\ref{T1DNear}) for non-Dirichlet boundary
conditions. For Dirichlet boundary condition the leading asymptotic is given
by the same expressions with the opposite sign. The leading terms vanish for
a conformally coupled field and in this case the divergences on the brane
are weaker. At large proper distances from the brane and for non-Neumann
boundary conditions, the diagonal components of the vacuum energy-momentum
tensor decay like $(z/x^{1})^{D+2\nu }$ and the off-diagonal component
behaves as $(z/x^{1})^{D+2\nu +1}$. In the case of Neumann boundary
condition the asymptotics have the opposite sign. For fixed $x^{1}$, the
diagonal components decay on the AdS boundary as $z^{D+2\nu }$ and diverge
on the horizon as $z^{D+1}$.

\section*{Acknowledgments}

E.R.B.M. and A.A.S. thank Conselho Nacional de Desenvolvimento Cient\'{\i}%
fico e Tecnol\'{o}gico (CNPq) for the financial support. A.A.S. was
supported by the State Committee of Science of the Ministry of Education and
Science RA, within the frame of Grant No. SCS 13-1C040.

\end{document}